\documentstyle[amssymb,aps,12pt]{revtex}

\begin{document}
\draft
\author{Sergio De Filippo}
\address{Tel: +39 089 965 229, Fax: +39 089 965 275, e-mail: defilippo@sa.infn.it}
\author{Filippo Maimone}
\address{maimone@sa.infn.it}
\address{Dipartimento di Fisica ''E.R. Caianiello'', Universit\`{a} di Salerno\\
Via Allende I-84081 Baronissi (SA) ITALY\\
and \\
Unit\`{a} I.N.F.M., I.N.F.N. Salerno}
\date{\today}
\title{Non-unitary HD gravity classically equivalent to Einstein gravity and its
Newtonian limit}
\maketitle

\begin{abstract}
Runaway solutions can be avoided in fourth order gravity by a doubling of
the matter operator algebra with a symmetry constraint with respect to the
exchange of observable and hidden degrees of freedom together with the
change in sign of the ghost and the dilaton fields. The theory is
classically equivalent to Einstein gravity, while its non-unitary Newtonian
limit is shown to lead to a sharp transition, around $10^{11}$ proton
masses, from the wavelike properties of microscopic particles to the
classical behavior of macroscopic bodies, as well as to a trans-Planckian
regularization of collapse singularities. A unified reading of ordinary and
black hole entropy emerges as entanglement entropy with hidden degrees of
freedom. The emergent picture gives a substantial agreement with B-H entropy
and Hawking temperature.
\end{abstract}


\bigskip Phys. Rev. D, to appear

\section{Introduction}

A possible way out of the so called information loss paradox\cite
{hawking1,preskill} emerging from black hole physics\cite{hawking2} consists
in assuming a fundamental non-unitarity\cite{ellis,banks,unruh,kay,wald}. In
fact it is natural to expect that the decoherence due to black hole
formation and evaporation should give rise to a significant modification of
the dynamical evolution laws: ''For almost any initial quantum state, one
would expect there to be a nonvanishing amplitude for black hole formation
and evaporation to occur - at least at a highly microscopic (e.g.,
Planckian) scale - thereby giving rise to a nonvanishing probability for
evolution from pure states to mixed states''\cite{unruh}. Although such an
evolution is incompatible with a cherished principle of quantum theory,
which postulates a unitary time evolution of a state vector in a Hilbert
space, the crucial issue is to assess if it necessarily gives rise to a loss
of quantum coherence or to violations of energy-momentum conservation so
large as to be incompatible with ordinary laboratory physics \cite
{ellis,banks,unruh,wald}. Arguments for such violations were given, starting
from the assumption that the effective evolution law governing laboratory
physics has a Markovian\ character \cite{ellis,banks}. On the contrary one
would expect that an effective evolution law modeling the process of black
hole formation and evaporation, far from being local in time, should retain
a long term ``memory''\cite{unruh,wald}. In particular the basic idea of the
non-Markovian models considered in Ref. \cite{unruh} is to have the given
system interacting with a ''hidden system'' with ''no energy of its own and
therefore... not... available as either a net source or a sink of energy''.

On the other hand a mechanism for large entropy production in gravitational
collapses should most naturally operate in the high curvature region, where
one may expect new physics to emerge, while connecting it with the event
horizon is somehow puzzling, as the physics on such a manifold has nothing
peculiar for a free falling observer. Of course a quantitative model of
Bekenstein-Hawking (B-H) entropy\cite{bekenstein}, along these lines, has to
refer to the collapsed matter and, in order to do that, it has to include a
mechanism for the elimination of the singularity. This does not mean that
one can not identify the entropy carried by Hawking radiation as coming from
the horizon within a local viewpoint: the entropy growth outside the
horizon, instead of being directly connected with an entropy produced by a
strongly non-unitary dynamics in the region close to the classical
singularity, is locally seen as a \ transformation of entanglement entropy
into von Neumann entropy. In fact the relative character of the degrees of
freedom involved in a given entanglement entropy is present even in flat
space-times, where it can be exhibited explicitly\cite{gingrich}. Of course,
in trying to pass from the region close to the horizon, where conventional
quantum field theory in curved space-times is expected to work as a good
approximation, to the region close to the classical singularity, we have to
pay a price. In the absence of a full theory of quantum gravity, we have to
rely on partially heuristic arguments and some guessing work, which we
intend to show can be carried out by rather natural assumptions.

In looking for a non-unitary theory avoiding the collapse singularity, we
are going to start from higher derivative (HD) gravity. From a purely
cosmological viewpoint it achieved great popularity since an inflationary
solution was obtained without invoking phase transitions in the very early
universe, from a field equation containing only geometric terms \cite
{starobinsky}. More recently a renewed attention towards HD gravity was
sparked by the appearance of HD gravitational terms in the low-energy
effective action of string theory and in the holographic renormalization
group, as well as by a growing interest in the study of brane worlds in HD
gravity\cite{nojiri}. However, although HD theories of gravity are natural
generalizations of Einstein gravity, already on the classical level they are
unstable for the presence of negative energy fields giving rise to runaway
solutions. On the quantum level, as to unitarity, a possible optimistic
conclusion is that ''the S-matrix will be nearly unitary\cite{dewitt}''\cite
{hawkinghertog}. The crucial obstacle in trying to define HD gravity as a
sound physical theory, namely the presence of ghosts, seems \ in fact to be
a strong indication, on one side, of non-unitarity and, on the other, of a
possible mechanism for avoiding singularities, thanks to short range
repulsive terms.

Here a specific non-unitary realization of HD gravity is shown to be
compatible with the wavelike properties of microscopic particles, as well as
with the assumption of a gravity-induced emergence of classicality\cite
{karolyhazy,hawking3,diosi,ghirardi,penrose,squires,ellis2,anandan,power},
and seems to give strong indications for the elimination of singularities on
a trans-Planckian scale. Parenthetically we are encouraged in our
extrapolations by the success of inflationary models, implicitly referring
to these scales\cite{brandenberger}. The present setting suggests that B-H
entropy may be identified with the von Neumann entropy of the collapsed
matter, or equivalently with the entanglement entropy between matter and
hidden degrees of freedom, both close to the smoothed singularity. This
viewpoint is corroborated by the attractive features of the Newtonian limit
of the model. In this attempt of evaluating relevant physical quantities by
an incomplete theory, and in particular by its Newtonian limit, we are
encouraged by well-known precedents, like the amazing quantitative agreement
between the analysis of John Mitchell in 1784 and the modern notion of a
black hole.

A bonus of the present non-unitary model is the possibility of a unified
notion, as von Neumann entropy, both for B-H and ordinary thermodynamic
entropy of closed systems. This is not irrelevant, as, ''...in order to gain
a better understanding of the degrees of freedom responsible for black hole
entropy, it will be necessary to achieve a deeper understanding of the
notion of entropy itself. Even in flat space-time, there is far from
universal agreement as to the meaning of entropy -- particularly in quantum
theory -- and as to the nature of the second law of thermodynamics''\cite
{wald}.

\section{Stable fourth order gravity}

Long ago deWitt\cite{dewitt} and Stelle\cite{stelle} analyzed the improved
ultraviolet behavior of HD gravity theories stemming from cancellations that
are ''analogous to the Pauli-Villars regularization of other field
theories'' \cite{stelle}. These cancellations are precisely due to the
presence of negative energy fields, which in their turn are the source of
instability: energy can flow from negative energy degrees of freedom to
positive energy ones and one can have runaway solutions.

In Ref.\cite{hawkinghertog} a remedy for the ghost problem, leading to a
non-unitary theory, was suggested by a suitable redefinition of the
euclidean path integral. In this paper we mean to propose an approach
directly in real space-time, thus avoiding analytic continuation, which
amounts to a very tricky operation outside the realm of a fixed flat
geometry. Like in Ref.\cite{hawkinghertog}, classical instability is cured
at the expense of unitarity. Before treating the physically relevant case,
we consider first a simpler fourth order theory for a scalar field $\phi $,
which has the same ghostly behavior as HD gravity\cite{hawkinghertog}. Its
action is 
\begin{equation}
S=\int d^{4}x\left[ -\frac{1}{2}\phi \square \left( \square -\mu ^{2}\right)
\phi -\lambda \phi ^{4}+\alpha \psi ^{\dagger }\psi \phi \right] +S_{mat}%
\left[ \psi ^{\dagger },\psi \right] ,
\end{equation}
with the inclusion of a matter action $S_{mat}$ and an interaction with
matter, where $\psi ^{\dagger }\psi $ is a shorthand notation for a
quadratic scalar expression in matter operators. Defining 
\begin{equation}
\phi _{1}=\frac{\left( \square -\mu ^{2}\right) \phi }{\mu },\;\phi _{2}=%
\frac{\square \phi }{\mu },\;  \label{transformation}
\end{equation}
the action can be rewritten as 
\begin{eqnarray}
&&S\left[ \phi _{1},\phi _{2},\psi ^{\dagger },\psi \right] =S_{mat}\left[
\psi ^{\dagger },\psi \right]  \nonumber \\
&&+\int d^{4}x\left[ \frac{1}{2}\phi _{1}\square \phi _{1}-\frac{1}{2}\phi
_{2}\left( \square -\mu ^{2}\right) \phi _{2}-\lambda \left( \frac{\phi
_{1}-\phi _{2}}{\mu }\right) ^{4}+\frac{\alpha }{\mu }\psi ^{\dagger }\psi
\left( \phi _{2}-\phi _{1}\right) \right] .  \label{secordertoy}
\end{eqnarray}
The action of $\phi _{2}$ has the wrong sign, which classically means that
the energy of the $\phi _{2}$ field is negative. If there were no
interaction terms, this negative energy wouldn't matter because each of the
fields would live in its own world and the positive and negative energy
worlds would not communicate with each other. However, if there are
interaction terms, like $\phi ^{4}$ or $\psi ^{\dagger }\psi \phi $, energy
can flow from negative to positive energy degrees of freedom, and one can
have runaway solutions, with the positive energy of $\phi _{1}$ and the
negative energy of $\phi _{2}$ both increasing exponentially in time \cite
{hawkinghertog}.

This toy model shares with HD gravity theories some of the mentioned
improvements on the ultraviolet behavior. In fact there is a complete
cancellation of all infinities coming from the $\psi ^{\dagger }\psi \phi $
interaction and corresponding to self-energy and vertex graphs \cite
{itzykson}, owing to the difference in sign between $\phi _{1}$ and $\phi
_{2}$ propagators. A key feature of the non-interacting theory ($\lambda
=\alpha =0$), making it classically viable, can be considered to be its
symmetry under the transformation $\phi _{2}\longrightarrow -\phi _{2}$, by
which symmetrical initial conditions, i.e. with $\phi _{2}=\dot{\phi}_{2}=0$%
, produce symmetrical solutions, thus in particular avoiding the runaway
ones. We are going now to extend this symmetry to the interacting theory. If
one symmetrizes the action (\ref{secordertoy}) as it is, this eliminates the
direct interaction between the ghost field and the matter altogether and
then the mentioned cancellations. A possible procedure to get a symmetric
action while keeping Pauli-Villars-like cancellations is suggested by
previous attempts \cite{hawkinghertog} and by the information loss paradox 
\cite{unruh}, both pointing to a non-unitary theory, where tracing out
hidden degrees of freedom results in general in mixed states. In particular
the most natural way to make the hidden degrees of freedom ''not...
available as either a net source or a sink of energy''\cite{unruh} is to
constraint them to be an exact copy of the observable ones. Accordingly we
introduce a (meta-)matter algebra that is the product of two equivalent
copies of the observable matter algebra, respectively generated by the $\psi
^{\dagger },\psi $ and $\tilde{\psi}^{\dagger },\tilde{\psi}$ operators, and
a symmetrized action 
\begin{equation}
S_{Sym}=\frac{1}{2}\left\{ S\left[ \phi _{1},\phi _{2},\psi ^{\dagger },\psi %
\right] +S\left[ \phi _{1},-\phi _{2},\tilde{\psi}^{\dagger },\tilde{\psi}%
\right] \right\} ,  \label{symac}
\end{equation}
which is invariant under the symmetry transformation 
\begin{equation}
\phi _{2}\longrightarrow -\phi _{2},\;\psi \longrightarrow \tilde{\psi},\;\;%
\tilde{\psi}\longrightarrow \psi \;.\;
\end{equation}

This duplication is formally analogous to what is done in thermo-field
dynamics\cite{umezawa}, where in particular it can be used to describe the
irreversible evolution of open systems\cite{arimitsu}. If the symmetry
constraint is imposed on quantum states, i.e. the state space is restricted
to those states $\left| \Psi \right\rangle $ that are generated from the
vacuum by symmetrical operators, then 
\begin{equation}
\left\langle \Psi \right| F\left[ \phi _{2},\psi ^{\dagger },\psi \right]
\left| \Psi \right\rangle =\left\langle \Psi \right| F\left[ -\phi _{2},%
\tilde{\psi}^{\dagger },\tilde{\psi}\right] \left| \Psi \right\rangle
\;\;\forall F.
\end{equation}
This implies that, as usual with constrained theories, allowed states do not
give a faithful representation of the original algebra, which is then larger
than the observable algebra. In particular the constrained state space
cannot distinguish between $F\left[ \psi ^{\dagger },\psi \right] $ and $F%
\left[ \tilde{\psi}^{\dagger },\tilde{\psi}\right] $, by which the $\tilde{%
\psi}$ operators are referred to hidden degrees of freedom, according to a
standard terminology for non-unitary models\cite{unruh}, while only the $%
\psi $ operators represent matter degrees of freedom. On a classical level
the constraint implies that $\psi $ and $\tilde{\psi}$ are to be identified
while the $\phi _{2}$ field vanishes and, as a consequence, the classical
constrained action is that of an ordinary second order scalar theory
interacting with matter: 
\begin{equation}
S_{Cl}=\int d^{4}x\left[ \frac{1}{2}\phi _{1}\square \phi _{1}-\lambda
\left( \frac{\phi _{1}}{\mu }\right) ^{4}-\frac{\alpha }{\mu }\phi _{1}\psi
^{\dagger }\psi \right] +S_{mat}\left[ \psi ^{\dagger },\psi \right] .
\end{equation}

Consider then the classical action for a fourth order theory of gravity
including matter\cite{stelle}

\begin{eqnarray}
S &=&S_{G}\left[ g_{\mu \nu }\right] +S_{mat}\left[ g_{\mu \nu },\psi
^{\dagger },\psi \right]  \nonumber \\
&=&-\int d^{4}x\sqrt{-g}\left[ \alpha {\cal R}_{\mu \nu }{\cal R}^{\mu \nu
}-\beta {\cal R}^{2}+\frac{1}{16\pi G}{\cal R}\right] +\int d^{4}x\sqrt{-g}%
L_{mat},  \label{hd}
\end{eqnarray}
where $L_{mat}$ denotes the matter Lagrangian density in a generally
invariant form. In terms of the contravariant metric density $\sqrt{32\pi G}%
h^{\mu \nu }=\sqrt{-g}g^{\mu \nu }-\eta ^{\mu \nu }$, the Newtonian limit of
the static field gives 
\begin{equation}
h^{00}\sim \frac{1}{r}+\frac{1}{3}\frac{e^{-\mu _{0}r}}{r}-\frac{4}{3}\frac{%
e^{-\mu _{2}r}}{r},  \label{potential}
\end{equation}
where $\mu _{0}=[32\pi G(3\beta -\alpha )]^{-1/2}$, $\mu _{2}=[16\pi G\alpha
]^{-1/2}$\cite{stelle}, while a complete analysis for the whole metric can
be found in Ref. \cite{schmidt}. From Stelle's linearized analysis, the
first term in Eq. (\ref{potential}) corresponds to the usual massless
graviton, the second one to a massive scalar and the third one to a negative
energy spin-two field. In fact, in analogy with Eq. (\ref{transformation}),
one can introduce an explicit transformation from the initial field $g_{\mu
\nu }$ appearing in the fourth order form of the action to a new metric
tensor $\bar{g}_{\mu \nu }$, a massless scalar field $\chi $ dilatonically
coupled to the metric and a spin-two massive field $\phi _{\mu \nu }$, this
transformation leading to the second order form\cite{hindawi}. To be
specific, following Ref.\cite{hindawi} (see Eq. (6.9) apart from the matter
term), the action (\ref{hd}) can be rewritten as the sum of the
Einstein-Hilbert action $S_{EH}$ for $\bar{g}_{\mu \nu }$, an action $S_{gh}$
for the traceless symmetric ghost field $\phi _{\mu \nu }$ and the scalar
field $\chi $ coupled to the metric $\bar{g}_{\mu \nu }$, and a matter
action $S_{mat}$, where $g_{\mu \nu }$ is expressed in terms of $\bar{g}%
_{\mu \nu }$, $\phi _{\mu \nu }$ and $\chi $ (replacing $g_{\mu \nu }$ by $%
e^{\chi }g_{\mu \nu }$ in Eq. (4.12) in Ref. \cite{hindawi}): 
\begin{eqnarray}
&&S\left[ \bar{g}_{\mu \nu },\phi _{\mu \nu },\chi ,\psi ^{\dagger },\psi %
\right]  \nonumber \\
&=&S_{EH}\left[ \bar{g}_{\mu \nu }\right] +S_{gh}\left[ \bar{g}_{\mu \nu
},\phi _{\mu \nu },\chi \right] +S_{mat}\left[ g_{\mu \nu }(\bar{g}_{\sigma
\tau },\phi _{\sigma \tau },\chi ),\psi ^{\dagger },\psi \right] .
\label{secorder}
\end{eqnarray}

In $S_{gh}$ above the quadratic part in $\phi _{\mu \nu }$ has the wrong
sign \cite{hindawi}, just as for $\phi _{2}$ in Eq. (\ref{secordertoy}). As
before, a simple way to get rid of classical instability would be to
symmetrize the action with respect to the transformation $\phi _{\mu \nu
}\rightarrow -\phi _{\mu \nu }$ and to introduce the symmetry constraint
with respect to this transformation. This however would eliminate the
corresponding repulsive term in Eq. (\ref{potential}), which is a possible
candidate in avoiding the singularity in gravitational collapse. Once one
accepts non-unitarity, it is rather natural to assume that one can cure the
instability, while keeping the short-range repulsive term, by introducing
hidden degrees of freedom as above, i.e. from a quantum viewpoint to accept
that the operator algebra involved in defining the dynamics is larger than
the observable algebra. To be specific, once again, we double the matter
algebra by taking a meta-matter algebra which is the product of two copies
of the observable matter algebra, respectively generated by the operators $%
\psi ^{\dagger },\psi $ and $\tilde{\psi}^{\dagger },\tilde{\psi}$. We then
define the symmetrized action\cite{defilippo5} 
\begin{equation}
S_{Sym}=\frac{1}{2}\left\{ S\left[ \bar{g}_{\mu \nu },\phi _{\mu \nu },\chi
,\psi ^{\dagger },\psi \right] +S\left[ \bar{g}_{\mu \nu },-\phi _{\mu \nu
},-\chi ,\tilde{\psi}^{\dagger },\tilde{\psi}\right] \right\} ,
\label{covsymac}
\end{equation}
which is symmetric under the transformation 
\begin{equation}
\phi _{\sigma \tau }\rightarrow -\phi _{\sigma \tau },\;\chi \longrightarrow
-\chi ,\;\psi \longrightarrow \tilde{\psi},\;\tilde{\psi}\longrightarrow
\psi ,\;\bar{g}_{\mu \nu }\longrightarrow \bar{g}_{\mu \nu }.\;
\end{equation}
Like above, if the state space is restricted to those states $\left| \Psi
\right\rangle $ that are generated from the vacuum by symmetrical operators,
then 
\begin{equation}
\left\langle \Psi \right| F\left[ \bar{g}_{\mu \nu },\phi _{\mu \nu },\chi
,\psi ^{\dagger },\psi \right] \left| \Psi \right\rangle =\left\langle \Psi
\right| F\left[ \bar{g}_{\mu \nu },-\phi _{\mu \nu },-\chi ,\tilde{\psi}%
^{\dagger },\tilde{\psi}\right] \left| \Psi \right\rangle \;\;\forall F.
\label{constraint}
\end{equation}
Like for the previous toy model, the constrained state space does not
distinguish between $F\left[ \psi ^{\dagger },\psi \right] $ and $F\left[ 
\tilde{\psi}^{\dagger },\tilde{\psi}\right] $, by which the $\tilde{\psi}$
operators are referred to hidden degrees of freedom, while only the $\psi $
operators represent observable matter. On a classical level the constraint
implies that $\psi $ and $\tilde{\psi}$ are to be identified, while the $%
\phi _{\mu \nu }$ and $\chi $ fields vanish and, as a consequence, the
classical constrained action is that of \ ordinary matter coupled to
ordinary gravity: 
\begin{equation}
S_{Cl}\left[ \bar{g}_{\mu \nu },\psi ^{\dagger },\psi \right] =S_{EH}\left[ 
\bar{g}_{\mu \nu }\right] +S_{mat}\left[ \bar{g}_{\mu \nu },\psi ^{\dagger
},\psi \right] ,
\end{equation}
as $S_{gh}\left[ \bar{g}_{\mu \nu },0,0\right] =0$ (Eq. (6.9) in Ref. \cite
{hindawi}) and $g_{\mu \nu }(\bar{g}_{\sigma \tau },0,0)=\bar{g}_{\mu \nu }$
(Eq. (4.12) in Ref. \cite{hindawi} with $e^{\chi }g_{\mu \nu }$ replacing $%
g_{\mu \nu }$). While this modification of fourth order gravity is expected
to affect its ultraviolet behavior, still it does not worsen it for one-loop
meta-matter to meta-matter amplitudes, at variance with the trivial
symmetrization of the original theory. It should also be remarked that one
could limit symmetrization to the ghost field only, without involving the
scalar field, especially if one were concerned with the cosmological
implications of keeping the dilatonic scalar field in the classical action.

A final remark is in order as to the possible rereading of the present model
as a bimetric HD theory where two worlds interact only by means of a
coupling between the metrics (of the fourth order formalism)\cite{damour}.
In fact the model can be defined by replacing $\bar{g}_{\mu \nu }$, $-\phi
_{\mu \nu }$, and $-\chi $ in the second term of the right hand side of Eq. (%
\ref{covsymac}) by three independent fields and adding an interaction,
including Lagrange multipliers, leading to the necessary identifications.
However we do not commit ourselves to the prevailing view pointing to
underlying higher dimensional theories, even though a natural setting could
appear to be an extension of the Randall-Sundrum model\cite{randall}, with
two positive tension flat branes separated by one intermediate negative
tension flat brane \cite{kogan}, where $\psi $ and $\tilde{\psi}$
meta-matters reside on distinct positive tension branes.

\section{Newtonian Limit and Gravitational Localization}

Of course the elimination of classical runaway solutions is only a first
step in assessing the consistency of the ensuing non-unitary theory. A
further natural step consists in studying its main implications for ordinary
laboratory physics. In order to do that, consider the Newtonian limit of
such a theory with non-relativistic meta-matter and instantaneous action at
a distance interactions. Looking at the signs in Eq. (\ref{covsymac}), we
see the following. The interactions due to the massless graviton field $\bar{%
g}_{\mu \nu }$ are always attractive, whereas those due to the scalar field $%
\chi $ are attractive but for the ones between observable and hidden
meta-matter; finally those due to the massive field $\phi _{\mu \nu }$ are
repulsive within observable and within hidden meta-matter, due to its
ghostly character (see the sign of the third term in Eq. (\ref{potential})),
and are otherwise attractive, since the ghostly character is offset by the
difference in sign in its coupling with observable and hidden meta-matter.
The corresponding (meta-)Hamiltonian operator is then\cite{defilippo5} 
\[
H_{G}=H_{0}[\psi ^{\dagger },\psi ]+H_{0}[\tilde{\psi}^{\dagger },\tilde{\psi%
}] 
\]
\begin{eqnarray}
&&-\frac{G}{2}\sum_{j,k}m_{j}m_{k}\int dxdy\frac{\psi _{j}^{\dagger }(x)\psi
_{j}(x)\tilde{\psi}_{k}^{\dagger }(y)\tilde{\psi}_{k}(y)}{|x-y|}\left( 1-%
\frac{1}{3}e^{-\mu _{0}|x-y|}+\frac{4}{3}e^{-\mu _{2}|x-y|}\right)  \nonumber
\\
&&-\frac{G}{4}\sum_{j,k}m_{j}m_{k}\int dxdy\frac{\psi _{j}^{\dagger }(x)\psi
_{j}(x)\psi _{k}^{\dagger }(y)\psi _{k}(y)}{|x-y|}\left( 1+\frac{1}{3}%
e^{-\mu _{0}|x-y|}-\frac{4}{3}e^{-\mu _{2}|x-y|}\right)  \nonumber \\
&&-\frac{G}{4}\sum_{j,k}m_{j}m_{k}\int dxdy\frac{\tilde{\psi}_{j}^{\dagger
}(x)\tilde{\psi}_{j}(x)\tilde{\psi}_{k}^{\dagger }(y)\tilde{\psi}_{k}(y)}{%
|x-y|}\left( 1+\frac{1}{3}e^{-\mu _{0}|x-y|}-\frac{4}{3}e^{-\mu
_{2}|x-y|}\right) ,  \label{newtonlimit}
\end{eqnarray}
acting on the product $F_{\psi }\otimes F_{\tilde{\psi}}$ of the Fock spaces
of the (non-relativistic counterparts of the) $\psi $ and $\tilde{\psi}$
operators. Here two couples of non-relativistic meta-matter operators $\psi
_{j}^{\dagger },\psi _{j}$ and $\tilde{\psi}_{j}^{\dagger },\tilde{\psi}_{j}$
\ appear for every particle species and spin component, while $m_{j}$ is the
mass of the $j$-th particle species and $H_{0}$ is the matter Hamiltonian in
the absence of gravity. The $\tilde{\psi}_{j}$ operator obeys the same
statistics as the corresponding operators $\psi _{j}$, while $[\psi ,\tilde{%
\psi}]$ $_{-}=[\psi ,\tilde{\psi}^{\dagger }]_{-}=0$. Though never appearing
in our formulae, the electromagnetic potential in the Coulomb gauge should
be included in the original degrees of freedom, even though, in the
non-relativistic setting, it is not involved in the gravitational
interaction.

With reference to Eq. (\ref{newtonlimit}), observe that the action at a
distance counterpart of the field-theoretic cancellations mentioned above is
the possibility of avoiding normal ordering in the last two terms. It would
correspond, in fact, to the subtraction of the finite operator $G(\mu
_{0}-4\mu _{2})\sum_{j}m_{j}^{2}\int dx\psi _{j}^{\dagger }(x)\psi
_{j}(x)/12 $ and its hidden correspondent, which in a fixed particle-number
space correspond to irrelevant finite constants.

To be specific, the meta-particle state space $S$ is the subspace of $%
F_{\psi }\otimes F_{\tilde{\psi}}$ including the meta-states obtained from
the vacuum $\left| \left| 0\right\rangle \right\rangle =\left|
0\right\rangle _{\psi }\otimes \left| 0\right\rangle _{\tilde{\psi}}$ by
applying operators built in terms of the products $\psi _{j}^{\dagger }(x)%
\tilde{\psi}_{j}^{\dagger }(y)$ and symmetrical with respect to the
interchange $\psi ^{\dagger }\leftrightarrow \tilde{\psi}^{\dagger }$,
which, then, have the same number of $\psi $ and $\tilde{\psi}$
meta-particles of each species. As the observable algebra is identified with
the $\psi $ operator algebra, expectation values can be evaluated by
preliminarily tracing out the $\tilde{\psi}$ operators. In particular, for
instance, the most general meta-state corresponding to one particle states
is represented by 
\begin{equation}
\left| \left| f\right\rangle \right\rangle =\int dx\int dyf(x,y)\psi
_{j}^{\dagger }(x)\tilde{\psi}_{j}^{\dagger }(y)\left| \left| 0\right\rangle
\right\rangle ,\;\;f(x,y)=f(y,x).
\end{equation}
This is a consistent definition since $H_{G}$\ generates a group of
(unitary) endomorphisms of $S$. A pure $n$-particle state, represented in
the traditional setting by 
\begin{equation}
\left| g\right\rangle \doteq \int d^{n}xg(x_{1},x_{2},...,x_{n})\psi
_{j_{1}}^{\dagger }(x_{1})\psi _{j_{2}}^{\dagger }(x_{2})...\psi
_{j_{n}}^{\dagger }(x_{n})\left| 0\right\rangle   \label{g}
\end{equation}
is represented in $S$ by the only meta-state that, by tracing out $\tilde{%
\psi}$ operators, gives the state $\left| g\right\rangle \left\langle
g\right| $, with $\left| g\right\rangle $ as in Eq. (\ref{g}), namely by 
\begin{equation}
\left| \left| g\otimes g\right\rangle \right\rangle \propto \int
d^{n}xd^{n}yg(x_{1},...,x_{n})g(y_{1},...,y_{n})\psi _{j_{1}}^{\dagger
}(x_{1})...\psi _{j_{n}}^{\dagger }(x_{n})\tilde{\psi}_{j_{1}}^{\dagger
}(y_{1})...\tilde{\psi}_{j_{n}}^{\dagger }(y_{n})\left| 0\right\rangle .
\label{initial}
\end{equation}

It should be remarked that, when our initial knowledge of the system state
is characterized by a density matrix, there is no unique prescription to
associate it with a pure meta-state. In such a case one has to consider the
possibility of using mixed meta-states to encode our incomplete knowledge.

Considering, for notational simplicity, particles of one and the same
species, the time derivative of the matter canonical momentum in a space
region $\Omega $ in the Heisenberg picture reads 
\[
\frac{d\overrightarrow{p}_{\Omega }}{dt}=-i\hslash \frac{d}{dt}\int_{\Omega
}dx\psi ^{\dagger }(x)\nabla \psi (x)\equiv \left. \frac{d\overrightarrow{p}%
_{\Omega }}{dt}\right| _{G=0}+\vec{F}_{G}=-\frac{i}{\hslash }\left[ 
\overrightarrow{p}_{\Omega },H_{0}[\psi ^{\dagger },\psi ]\right] 
\]
\begin{eqnarray}
&&+\frac{G}{2}m^{2}\int_{\Omega }dx\psi ^{\dagger }(x)\psi (x)\nabla
_{x}\int_{R^{3}}dy\frac{\tilde{\psi}^{\dagger }(y)\tilde{\psi}(y)}{\left|
x-y\right| }\left( 1-\frac{1}{3}e^{-\mu _{0}|x-y|}+\frac{4}{3}e^{-\mu
_{2}|x-y|}\right)   \nonumber \\
&&+\frac{G}{2}m^{2}\int_{\Omega }dx\psi ^{\dagger }(x)\psi (x)\nabla
_{x}\int_{R^{3}}dy\frac{\psi ^{\dagger }(y)\psi (y)}{\left| x-y\right| }%
\left( 1+\frac{1}{3}e^{-\mu _{0}|x-y|}-\frac{4}{3}e^{-\mu _{2}|x-y|}\right) .
\end{eqnarray}
The expectation of the gravitational force can be written as 
\[
\left\langle \vec{F}_{G}\right\rangle =
\]
\begin{eqnarray}
&&\frac{G}{2}m^{2}\left\langle \int_{\Omega }dx\psi ^{\dagger }(x)\psi
(x)\nabla _{x}\int_{\Omega }dy\frac{\tilde{\psi}^{\dagger }(y)\tilde{\psi}(y)%
}{\left| x-y\right| }\left( 1-\frac{1}{3}e^{-\mu _{0}|x-y|}+\frac{4}{3}%
e^{-\mu _{2}|x-y|}\right) \right\rangle   \nonumber \\
&&+\frac{G}{2}m^{2}\left\langle \int_{\Omega }dx\psi ^{\dagger }(x)\psi
(x)\nabla _{x}\int_{R^{3}\backslash \Omega }dy\frac{\tilde{\psi}^{\dagger
}(y)\tilde{\psi}(y)}{\left| x-y\right| }\left( 1-\frac{1}{3}e^{-\mu
_{0}|x-y|}+\frac{4}{3}e^{-\mu _{2}|x-y|}\right) \right\rangle   \nonumber \\
&&+\frac{G}{2}m^{2}\left\langle \int_{\Omega }dx\psi ^{\dagger }(x)\psi
(x)\nabla _{x}\int_{\Omega }dy\frac{\psi ^{\dagger }(y)\psi (y)}{\left|
x-y\right| }\left( 1+\frac{1}{3}e^{-\mu _{0}|x-y|}-\frac{4}{3}e^{-\mu
_{2}|x-y|}\right) \right\rangle   \nonumber \\
&&+\frac{G}{2}m^{2}\left\langle \int_{\Omega }dx\psi ^{\dagger }(x)\psi
(x)\nabla _{x}\int_{R^{3}\backslash \Omega }dy\frac{\psi ^{\dagger }(y)\psi
(y)}{\left| x-y\right| }\left( 1+\frac{1}{3}e^{-\mu _{0}|x-y|}-\frac{4}{3}%
e^{-\mu _{2}|x-y|}\right) \right\rangle ,
\end{eqnarray}
where, on allowed states, the first term vanishes for the antisymmetry of
the kernel $\nabla _{x}\left[ \left( 1-e^{-\mu _{0}|x-y|}/3+4e^{-\mu
_{2}|x-y|}/3\right) /\left| x-y\right| \right] $ and the symmetry constraint
on the state, while the third one vanishes, as it should be for
self-gravitating matter, just as a consequence of the antisymmetry of the
corresponding kernel. As is usual with the evaluation of forces between
macroscopic bodies, we can then approximate $\left\langle \psi ^{\dagger
}(x)\psi (x)\tilde{\psi}^{\dagger }(y)\tilde{\psi}(y)\right\rangle $ and $%
\left\langle \psi ^{\dagger }(x)\psi (x)\psi ^{\dagger }(y)\psi
(y)\right\rangle $ respectively by $\left\langle \psi ^{\dagger }(x)\psi
(x)\right\rangle \left\langle \tilde{\psi}^{\dagger }(y)\tilde{\psi}%
(y)\right\rangle $ and $\left\langle \psi ^{\dagger }(x)\psi
(x)\right\rangle \left\langle \psi ^{\dagger }(y)\psi (y)\right\rangle $ ,
as $x\in \Omega $ and $y\in R^{3}\backslash \Omega $. Finally, as $%
\left\langle \tilde{\psi}^{\dagger }(y)\tilde{\psi}(y)\right\rangle
=\left\langle \psi ^{\dagger }(y)\psi (y)\right\rangle $, we get 
\begin{equation}
\left\langle \vec{F}_{G}\right\rangle \simeq Gm^{2}\int_{\Omega
}dx\left\langle \psi ^{\dagger }(x)\psi (x)\right\rangle \nabla
_{x}\int_{R^{3}\backslash \Omega }dy\frac{\left\langle \psi ^{\dagger
}(y)\psi (y)\right\rangle }{\left| x-y\right| },
\end{equation}
namely the classical aspects of the interaction are the same as for the
traditional Newton interaction, consistently with the classical equivalence
of the original theory to Einstein gravity\cite{defilippo5}.

Although we are using the general Newtonian limit (\ref{newtonlimit}), it is
worthwhile to remark that we are mainly interested to two opposite
specialized limits.

The ordinary Newtonian limit, for ordinary laboratory physics, corresponds
to taking $\mu _{0},\mu _{2}\rightarrow \infty $, if $\mu _{0}^{-1}$ and $%
\mu _{2}^{-1}$ are assumed, as usual, of the order of the Planck length, in
which case the meta-Hamiltonian $H_{G}$ can be rewritten in the form 
\begin{equation}
H_{G}=H[\psi ^{\dagger },\psi ]+H[\tilde{\psi}^{\dagger },\tilde{\psi}]-%
\frac{G}{2}\sum_{j,k}m_{j}m_{k}\int dxdy\frac{\psi _{j}^{\dagger }(x)\psi
_{j}(x)\tilde{\psi}_{k}^{\dagger }(y)\tilde{\psi}_{k}(y)}{|x-y|},
\label{muinfinity}
\end{equation}
where $H[\psi ^{\dagger },\psi ]$ and $H[\tilde{\psi}^{\dagger },\tilde{\psi}%
]$\ respectively include the halved (normal ordered) Newton interaction
within observable and hidden meta-matter. In this form we have a well
defined non-unitary model of Newtonian gravity without any free parameter.
Tracing out the $\tilde{\psi}$ operators from the meta-state evolving
according to the unitary meta-dynamics generated by $H_{G}$ results in a
non-Markov non-unitary physical dynamics for the ordinary matter algebra\cite
{defilippo1}.

The trans-Planckian Newtonian limit concerns the use we are going to make of
the model with reference to gravitational collapse, where the model replaces
the classical singularity with a trans-Planckian structure. To this end we
consider the opposite limit $\mu _{0},\mu _{2}\rightarrow 0$, leading to a
Hamiltonian $H_{G}$ as in (\ref{muinfinity}) with $H$ and $G/2$ respectively
replaced by $H_{0}$ and $G$. The rationale for the use of this bold
extension of the Newtonian limit outside its typical applicability range,
though within a merely heuristic approach, resides in part in the soundness
of its physical consequences, as shown in the following.

A general new feature of the model with respect to the usual inclusion of
Newtonian gravity in QM is the localization due to the presence of an
effective self-interaction. Consider in fact in the traditional setting a
physical body in a given quantum state whose wave function $\Psi
_{CM}(X)\Psi _{INT}(x_{i}-x_{j})$ is the product of the wave function of the
center of mass and of an internal wave function. In particular $\Psi _{CM}$
can be chosen, for simplicity, in such a way that the corresponding
meta-wave function $\Psi _{TOT}=\Psi _{CM}(X)\Psi _{INT}(x_{i}-x_{j})\Psi
_{CM}(Y)\Psi _{INT}(y_{i}-y_{j})$ can be rewritten as: 
\begin{equation}
\Psi _{TOT}=\tilde{\Psi}_{CM}(\frac{X+Y}{2})\tilde{\Psi}_{INT}(X-Y)\Psi
_{INT}(x_{i}-x_{j})\Psi _{INT}(y_{i}-y_{j}),  \label{localized}
\end{equation}
where $y_{i},Y$ denote the hidden correspondents of $x_{i},X$. As to $\tilde{%
\Psi}_{INT}(X-Y)$, we choose it as the ground state of the relative motion
of the two interpenetrating meta-bodies, which is formally equivalent to the
plasma oscillations of two opposite charge distributions. The corresponding
potential energy, if the body is spherically symmetric and not too far from
being a homogeneous distribution of radius $\Xi $ and mass $M$, has the form 
$\xi GM^{2}f\left( \left| X-Y\right| \right) $, where

\begin{equation}
f\left( r\right) =\left\{ 
\begin{array}{c}
-1/r\;\;\;\;\text{for }r\geq 2\Xi \\ 
\frac{1}{2}\alpha r^{2}/\Xi ^{3}\text{\ \ \ \ for }r\ll \Xi
\end{array}
\right. ,
\end{equation}
with $\xi =1/2,1$ respectively for the ordinary and the trans-Planckian
limit, and $\alpha \sim $ $10^{0}$ a dimensionless constant. We are
interested here to the case of small relative displacements. The relative
ground state is represented by 
\begin{equation}
\tilde{\Psi}_{INT}(X-Y)=\left( \Lambda ^{2}\pi \right) ^{-3/4}e^{\frac{%
-\left| X-Y\right| ^{2}}{2\Lambda ^{2}}};\;\ \ \;\Lambda =(2\hslash ^{2}\Xi
^{3}/\alpha \xi GM^{3})^{1/4}.  \label{gaussian}
\end{equation}
Then, if we choose $\Psi _{CM}(X)\propto \exp \left[ -X^{2}/\Lambda ^{2}%
\right] $, we get 
\begin{equation}
\tilde{\Psi}_{0}(X,Y)\equiv \Psi _{CM}(X)\Psi _{CM}(Y)=\tilde{\Psi}%
_{INT}(X-Y)\tilde{\Psi}_{INT}(X+Y).  \label{unentangled}
\end{equation}
In particular for body densities $\sim 10^{24}m_{p}/cm^{3}$, where $m_{p}$
denotes the proton mass, $\Lambda \sim (m_{p}/M)^{1/2}cm$, which shows that
the small displacement approximation is acceptable already for $M\sim
10^{12}m_{p}$, when $\Lambda \sim 10^{-6}cm$, whereas the body dimensions
are $\sim 10^{-4}cm$\cite{defilippo1}.

Another simple case corresponds to masses lower than $10^{10}m_{p}$, where
the two meta-bodies can be approximated as point particles and their ground
state wave function, in the ordinary Newtonian limit, is 
\begin{equation}
\Psi (X-Y)\propto e^{-\left| X-Y\right| /a};\;\;\;a=4\hslash ^{2}\xi
^{-1}G^{-1}M^{-3}\sim 10^{25}\left( M/m_{p}\right) ^{-3}cm,
\end{equation}
by which gravitational localization, consistently with recent experiments,
can be ignored for all practical purposes even for particles much larger
than fullerene\cite{rae,arndt}. The ensuing situation corresponds then to a
rather sharp localization mass threshold $M_{t}\sim \hslash
^{3/5}G^{-3/10}\rho ^{1/10}$, which is very robust with respect to mass
density variation.

It is easily seen that the present framework actually is compatible with the
way terrestrial gravity appears in QM. A crucial experiment, dating back to
1975, exhibits in fact in a striking manner how terrestrial gravity enters
the Schr\"{o}dinger equation in the usual way, i.e. just as a Coulomb
external field \cite{colella}. To this end the calculation of the average
gravitational force acting over a lump performed above does not suffice
since it can explain only e.g. the free fall of a microscopic particle by
means of classical equations (Ehrenfest theorem) where $\hbar $ does not
appear.

Consider the problem of a large, for simplicity spherically symmetric,
massive body (the Earth) in some irrelevant internal state in interaction
with an external microscopic particle. Define the meta-Hamiltonian of the
Earth-particle system as 
\begin{equation}
H=\frac{-\hbar ^{2}}{2M}\sum_{i=1,2}\nabla _{R_{i}}^{2}-GM^{2}f\left( \left|
R_{1}-R_{2}\right| \right) -\frac{\hbar ^{2}}{2m}\sum_{i=1,2}\nabla
_{x_{i}}^{2}-GmM\sum_{i,j=1,2}\frac{1}{\left| x_{i}-R_{j}\right| }
\end{equation}
where $M$ and $m$ are respectively the mass of the Earth and of the
particle, $R_{1},R_{2}$ and $x_{1},x_{2}$ respectively the center of
meta-mass coordinates of the two Earth and particle copies.

Let's start with a meta-state of the meta-Earth system corresponding to the
fundamental (or a not too highly excited ) one with respect to the relative
motion of the two copies and choosing the initial CM meta-state of the bound
system of the two copies just as above. The localization length is in this
case of the order $\Lambda ^{Earth}\sim $ $10^{-26}%
\mathop{\rm cm}%
$. Having in mind that the particle is described by a wave packet whose size 
$a$ is in any case much larger than $\Lambda ^{Earth}$, we can approximate
the squared modulus of Earth's meta-wave-function by a product of delta
functions $\delta ^{3}\left( r\right) \delta ^{3}\left( R\right) $, where $r$%
, $R$ respectively denote the internal and the CM coordinates of the
meta-Earth bound system. As a consequence $x_{i}-R_{j}$ in the Newton
potential can be replaced by $x_{i}$. Of course, since the spreading time of
the\ Earth's CM wave function over a region of the size $a\gtrsim 10^{-10}$ $%
\mathop{\rm cm}%
$ is given by $a\Lambda ^{Earth}M/\hbar \gtrsim 10^{19}\;%
\mathop{\rm s}%
$, the approximation is justified in any physically relevant situation, and
actually even much better than what appears from this analysis, as we are
ignoring the spreading of the particle wave function. As a result the
gravitational interaction enters in the particle dynamics simply by the
presence of the usual external Newton potential.

\section{Evolution from pure to mixed states}

It should be stressed that, while in the ensuing dynamics the constraint on
the hidden degrees of freedom to have the same average energy as the
observable matter avoids them to be ''available as either a net source or a
sink of energy'', only the meta-Hamiltonian is strictly conserved. If we
include in the physical energy the usual Newtonian interaction between
observable degrees of freedom, the physical energy operator 
\begin{equation}
H_{Ph}[\psi ^{\dagger },\psi ]=H_{0}[\psi ^{\dagger },\psi ]-\frac{G}{2}%
\sum_{j,k}m_{j}m_{k}\int dxdy\frac{:\psi _{j}^{\dagger }(x)\psi _{j}(x)\psi
_{k}^{\dagger }(y)\psi _{k}(y):}{|x-y|}
\end{equation}
is not the generator of time evolution. To be specific, in the ordinary
Newtonian limit, the generator of the meta-dynamics can be written 
\begin{eqnarray}
H_{G} &=&H_{Ph}[\psi ^{\dagger },\psi ]+H_{Ph}[\tilde{\psi}^{\dagger },%
\tilde{\psi}] \\
&&-\frac{G}{4}\sum_{j,k}m_{j}m_{k}\int dxdy\left[ \frac{2\psi _{j}^{\dagger
}(x)\psi _{j}(x)\tilde{\psi}_{k}^{\dagger }(y)\tilde{\psi}_{k}(y)}{|x-y|}%
\right]  \nonumber \\
&&+\frac{G}{4}\sum_{j,k}m_{j}m_{k}\int dxdy\left[ \frac{:\psi _{j}^{\dagger
}(x)\psi _{j}(x)\psi _{k}^{\dagger }(y)\psi _{k}(y):+:\tilde{\psi}%
_{j}^{\dagger }(x)\tilde{\psi}_{j}(x)\tilde{\psi}_{k}^{\dagger }(y)\tilde{%
\psi}_{k}(y):}{|x-y|}\right] ,  \nonumber
\end{eqnarray}
from which we see that $H_{G}$ and $H_{Ph}[\psi ^{\dagger },\psi ]+H_{Ph}[%
\tilde{\psi}^{\dagger },\tilde{\psi}]$ in general are different only due to
correlations. The two sums above have approximately equal expectations and
fluctuate around the classical gravitational energy. On one side these
energy fluctuations have to be present in any model leading to dynamical
wave function localization, which in itself requires a certain injection of
energy\cite{squires1}. On the other hand these fluctuations, though
irrelevant on a macroscopic scale, are precisely what can lead to
thermodynamical equilibrium in a closed system if thermodynamic entropy is
identified with von Neumann entropy\cite{kay}. In fact, due to the
interaction with the hidden degrees of freedom, a pure eigenstate of the
ordinary energy $H_{Ph}$\ is expected to evolve into a microcanonical
ensemble.

As a simple example showing how a pure state can evolve into a mixed one,
consider a free spherically symmetric body of ordinary matter above
localization threshold, initially described by a gaussian wave packet, whose
size is chosen as above in such a way that the particle-copy system is in
its ground state, thus recovering the meta-wave-function (\ref{localized}) 
\cite{defilippo2}.

The factor depending on the center of meta-mass of the $\psi $ and $\tilde{%
\psi}$ meta-bodies in $\tilde{\Psi}_{0}(X,Y)$\ (\ref{unentangled}), for $%
M\gtrsim 10^{12}m_{p}$, spreads in time as usual for a body of mass $2M$, so
that after a time $t$, the meta-wavefunction becomes 
\begin{equation}
\tilde{\Psi}_{t}(X,Y)\propto \exp \left[ \frac{-\left| X-Y\right| ^{2}}{%
2\Lambda ^{2}}\right] \exp \left[ \frac{-\left| X+Y\right| ^{2}/4}{\Lambda
^{2}/2+i\hslash t/M}\right] \equiv e^{-\alpha _{0}\left| X-Y\right|
^{2}}e^{-\alpha _{t}\left| X+Y\right| ^{2}}.  \label{evolvedgaussian}
\end{equation}
In order that this be compatible with the assumption that gravity
continuously forces localization\cite
{karolyhazy,hawking3,diosi,ghirardi,penrose,squires,ellis2,anandan,power},
the spreading of the physical state must be the outcome of the entropy
growth. This initially vanishes, as the initial meta-wavefunction (\ref
{unentangled}) is unentangled and then the physical state, obtained by
tracing out $Y$, is pure. If one evaluates the physical state $\rho
_{t}(X,X^{\prime })=\int dY\tilde{\Psi}_{t}(X,Y)\tilde{\Psi}_{t}^{\ast
}(X^{\prime },Y)$, one finds that the space probability density reads 
\begin{equation}
\rho _{t}(X,X)=\left[ \frac{8\alpha _{0}(\alpha _{t}+\bar{\alpha}_{t})}{\pi
(\alpha _{t}+\bar{\alpha}_{t}+2\alpha _{0})}\right] ^{3/2}\exp \left[ -\frac{%
8\alpha _{0}(\alpha _{t}+\bar{\alpha}_{t})}{(\alpha _{t}+\bar{\alpha}%
_{t}+2\alpha _{0})}X^{2}\right] \propto \exp \frac{-2\Lambda ^{2}X^{2}}{%
\Lambda ^{4}+2\hslash ^{2}t^{2}/M^{2}}.  \label{probability}
\end{equation}
The spreading is extremely slow, as its typical time, for bodies of density $%
\sim 10^{24}m_{p}/cm^{3}$, is $\sim 10^{3}\sec $ independently from the
mass, as can be checked by means of Eqs. (\ref{probability}) and (\ref
{gaussian}). If it is due to entropy growth only, rather than to the
spreading of the wave function, the entropy $S_{t}$ is expected to depend
approximately on the ratio between the final and the initial space volumes
roughly occupied by the two Gaussian densities, according to 
\begin{equation}
S_{t}\sim K_{B}\frac{3}{2}\ln \left[ \frac{\alpha _{t}+\bar{\alpha}%
_{t}+2\alpha _{0}}{2(\alpha _{t}+\bar{\alpha}_{t})}\right] ,
\label{approxentropy}
\end{equation}
at least for large enough times. (Linear momentum probability density does
not depend on time.) Of course this corresponds to the approximation of the
mixed state by means of an ensemble of $N$ equiprobable localized states,
which is legitimate if $N$ turns out to be large enough. In order to
evaluate the entropy of the state represented by $\rho _{t}(X,X^{\prime })$
and to check Eq. (\ref{approxentropy}), we use the possibility, in this
approximation, of linking the entropy 
\begin{equation}
S_{t}=-K_{B}\text{ }Tr\left[ \rho _{t}\ln \rho _{t}\right] =K_{B}\ln N
\label{equientropy}
\end{equation}
with the purity 
\begin{equation}
Tr\left[ \rho _{t}^{2}\right] =1/N;\;\;\;\;\;\;\rho _{t}^{2}(X,X^{\prime
})=\int dX^{\prime \prime }\rho _{t}(X,X^{\prime \prime })\rho
_{t}(X^{\prime \prime },X^{\prime }).  \label{equipurity}
\end{equation}
By an explicit computation we get 
\begin{equation}
Tr\left[ \rho _{t}^{2}\right] =\int dX\rho _{t}^{2}(X,X)=\frac{\left[
4\alpha _{0}(\alpha _{t}+\bar{\alpha}_{t})\right] ^{3}}{\left[ \left(
2\alpha _{t}\bar{\alpha}_{t}+6\alpha _{t}\alpha _{0}+6\bar{\alpha}_{t}\alpha
_{0}+2\alpha _{0}^{2}\right) ^{2}-4\left( \bar{\alpha}_{t}-\alpha
_{0}\right) ^{2}\left( \alpha _{t}-\alpha _{0}\right) ^{2}\right] ^{3/2}}
\end{equation}
and, for large times, one can keep just the leading term in $\alpha
_{t}/\alpha _{0}$, that is 
\begin{equation}
Tr\left[ \rho _{t}^{2}\right] \sim \left( \frac{\alpha _{t}+\bar{\alpha}_{t}%
}{2\alpha _{0}}\right) ^{3/2}
\end{equation}
which, by using Eqs. (\ref{equientropy},\ref{equipurity}), gives 
\begin{equation}
S_{t}\sim -K_{B}\frac{3}{2}\ln \left( \frac{\alpha _{t}+\bar{\alpha}_{t}}{%
2\alpha _{0}}\right) =K_{B}\frac{3}{2}\ln \left( \frac{\Lambda ^{4}+4\hslash
^{2}t^{2}/M^{2}}{\Lambda ^{4}}\right)
\end{equation}
which differs from the leading term in Eq.(\ref{approxentropy}) by an
irrelevant quantity $(3/2)K_{B}\ln 2$. This validates our view of the free
motion of a macroscopic body, at variance with the rather unphysical
stationary localized states of the Schr\"{o}dinger-Newton (S-N) model, whose
initial linear momentum uncertainty does not give rise to a spreading of the
probability density\cite{christian,penrose1,moroz,kumar,melko}. More
generally, while that non-linear generalization of QM was considered to be a
reasonable mean field approximation of an unspecified theory, by its
unitarity it can not model any fundamental gravitational decoherence. It is
remarkable that the S-N model can be actually obtained as the $N\rightarrow
\infty $ limit of the $N$ color generalization of the present Newtonian
limit \cite{defilippo4}.

It should be stressed that the notion of coarse graining entropy, often
taken as the starting point in dealing with the quantum foundations of the
second law of thermodynamics\cite{vonneumann}, can be easily connected with
the present approach. Consider, for simplicity, a non-degenerate physical
state 
\begin{equation}
\rho _{Ph}=\sum_{j}p_{j}\left| j\right\rangle \left\langle j\right|
,\;p_{j}\in R,\;\;p_{j}=p_{k}\Rightarrow j=k.\;
\end{equation}
The most general pure meta-state vector giving rise to $\rho _{ph}$ is 
\begin{equation}
\left| \left| \Psi \right\rangle \right\rangle _{{\bf \varphi }%
}=\sum_{j}e^{i\varphi _{j}}\sqrt{p_{j}}\left| j\right\rangle \left|
j\right\rangle ,  \label{microstate}
\end{equation}
where $\left| j\right\rangle \left| j\right\rangle $ denotes the tensor
product of two corresponding vectors in the two Fock spaces and the $\varphi
_{j}\in \lbrack 0,2\pi \lbrack $ are arbitrary real parameters. The
indistinguishability of the corresponding meta-states, due to the
restriction of the physical algebra, induces in the meta-state space an
unambiguous coarse graining, at variance with the rather vague one in the
traditional approaches. To be specific, it is natural to introduce the
macro-meta-state 
\begin{equation}
\rho _{CG}\equiv \int \prod_{j}\frac{d\varphi _{j}}{2\pi }\left| \left| \Psi
\right\rangle \right\rangle _{{\bf \varphi }}\left\langle \left\langle \Psi
\right| \right| =\sum_{j}p_{j}\left| j\right\rangle \left| j\right\rangle
\left\langle j\right| \left\langle j\right| ,
\end{equation}
corresponding to the equiprobability of the micro-meta-states $\left| \left|
\Psi \right\rangle \right\rangle _{{\bf \varphi }}\left\langle \left\langle
\Psi \right| \right| .$ The corresponding coarse graining entropy is 
\begin{equation}
S_{CG}=-K_{B}Tr\left[ \rho _{CG}\ln \rho _{CG}\right] =-K_{B}\sum_{j}p_{j}%
\ln p_{j},
\end{equation}
which coincides with the von Neumann entropy of the physical state $\rho
_{Ph}$.

Vice versa, if we assume that a specific pure meta-state $\left| \left| \Psi
\right\rangle \right\rangle $ is given, the Schmidt decomposition theorem
allows us to write it in terms of orthonormal vectors as 
\begin{equation}
\left| \left| \Psi \right\rangle \right\rangle =\sum_{j}\sqrt{p_{j}}\left|
j\right\rangle \left| j^{\prime }\right\rangle ,
\end{equation}
with the $p_{j}$ positive, for simplicity distinct, real numbers. By the
symmetry constraint on the meta-state space one can choose the relative
phases in such a way that $\left| j\right\rangle $ and $\left| j^{\prime
}\right\rangle $ can be taken as corresponding vectors in the two Fock
spaces, thus reproducing $\left| \left| \Psi \right\rangle \right\rangle _{%
{\bf \varphi }}$ in eq. (\ref{microstate}) for ${\bf \varphi }=0$. Although
this amounts to the knowledge of a definite microstate, the entropy of the
corresponding physical state $\rho _{Ph}$ is non-vanishing and coincides
with the coarse graining entropy of the corresponding macrostate $\rho _{CG}$%
. This shows the objective and non-conventional character of the notion of
entropy in the present approach, since it does not depend on a subjective
characterization based on the notion of a macroscopic observer \cite
{vonneumann}.

\section{Wave function reduction}

In an interaction representation of the ordinary Newtonian limit, where the
free meta-Hamiltonian is $H[\psi ^{\dagger },\psi ]+H[\tilde{\psi}^{\dagger
},\tilde{\psi}]$, the time evolution of an initially untangled meta-state $%
\left| \left| \tilde{\Phi}(0)\right\rangle \right\rangle $\ is represented
by 
\begin{eqnarray}
\left| \left| \tilde{\Phi}(t)\right\rangle \right\rangle &=&{\it T}\exp %
\left[ \frac{i}{\hslash }Gm^{2}\int dt\int dxdy\frac{\psi ^{\dagger
}(x,t)\psi (x,t)\tilde{\psi}^{\dagger }(y,t)\tilde{\psi}(y,t)}{|x-y|}\right]
\left| \left| \tilde{\Phi}(0)\right\rangle \right\rangle  \nonumber \\
&\equiv &U(t)\left| \left| \tilde{\Phi}(0)\right\rangle \right\rangle \equiv
U(t)\left| \Phi (0)\right\rangle _{\psi }\otimes \left| \Phi
(0)\right\rangle _{\chi }.  \label{evolvedmetastate}
\end{eqnarray}
Then, by a Stratonovich-Hubbard transformation\cite{negele}, we can rewrite $%
U(t)$ as 
\begin{eqnarray}
U(t) &=&\int {\it D}\left[ \varphi _{1},\varphi _{2}\right] \exp \frac{ic^{2}%
}{2\hslash }\int dtdx\left[ \varphi _{1}\nabla ^{2}\varphi _{1}-\varphi
_{2}\nabla ^{2}\varphi _{2}\right]  \nonumber \\
&&{\it T}\exp \left[ -i\frac{mc}{\hslash }\sqrt{2\pi G}\int dtdx\left[
\varphi _{1}(x,t)+\varphi _{2}(x,t)\right] \psi ^{\dagger }(x,t)\psi (x,t)%
\right]  \nonumber \\
&&{\it T}\exp \left[ -i\frac{mc}{\hslash }\sqrt{2\pi G}\int dtdx\left[
\varphi _{1}(x,t)-\varphi _{2}(x,t)\right] \tilde{\psi}^{\dagger }(x,t)%
\tilde{\psi}(x,t)\right]  \label{stratonovich}
\end{eqnarray}
namely as a functional integral over two auxiliary real scalar fields $%
\varphi _{1}$ and $\varphi _{2}$.

The physical state corresponding to the meta-state (\ref{evolvedmetastate})
is given by 
\begin{equation}
\rho _{Ph}(t)\equiv Tr_{\tilde{\psi}}\left| \left| \tilde{\Phi}%
(t)\right\rangle \right\rangle \left\langle \left\langle \tilde{\Phi}%
(t)\right| \right| =\sum_{k}\;\;\;_{\tilde{\psi}}\left\langle k\right|
\left| \left| \tilde{\Phi}(t)\right\rangle \right\rangle \left\langle
\left\langle \tilde{\Phi}(t)\right| \right| \left| k\right\rangle _{\tilde{%
\psi}}
\end{equation}
and, by using Eq. (\ref{stratonovich}), we can write 
\[
_{\tilde{\psi}}\left\langle k\right| \left| \left| \tilde{\Phi}%
(t)\right\rangle \right\rangle =\int {\it D}\left[ \varphi _{1},\varphi _{2}%
\right] \exp \frac{ic^{2}}{2\hslash }\int dtdx\left[ \varphi _{1}\nabla
^{2}\varphi _{1}-\varphi _{2}\nabla ^{2}\varphi _{2}\right] 
\]
\begin{eqnarray}
&&_{\tilde{\psi}}\left\langle k\right| {\it T}\exp \left[ -i\frac{mc}{%
\hslash }\sqrt{2\pi G}\int dtdx\left[ \varphi _{1}(x,t)-\varphi _{2}(x,t)%
\right] \tilde{\psi}^{\dagger }(x,t)\tilde{\psi}(x,t)\right] \left| \Phi
(0)\right\rangle _{\tilde{\psi}}  \nonumber \\
&&{\it T}\exp \left[ -i\frac{mc}{\hslash }\sqrt{2\pi G}\int dtdx\left[
\varphi _{1}(x,t)+\varphi _{2}(x,t)\right] \psi ^{\dagger }(x,t)\psi (x,t)%
\right] \left| \Phi (0)\right\rangle _{\tilde{\psi}}.
\end{eqnarray}
Then the final expression for the physical state at time $t$ is given by 
\[
\rho _{Ph}(t)=\int {\it D}\left[ \varphi _{1},\varphi _{2},\varphi
_{1}^{\prime },\varphi _{2}^{\prime }\right] \exp \frac{ic^{2}}{2\hslash }%
\int dtdx\left[ \varphi _{1}\nabla ^{2}\varphi _{1}-\varphi _{2}\nabla
^{2}\varphi _{2}-\varphi _{1}^{\prime }\nabla ^{2}\varphi _{1}^{\prime
}+\varphi _{2}^{\prime }\nabla ^{2}\varphi _{2}^{\prime }\right] 
\]
\begin{eqnarray}
&&_{\psi }\left\langle \Phi (0)\right| {\it T}^{-1}\exp \left[ i\frac{mc}{%
\hslash }\sqrt{2\pi G}\int dtdx\left[ \varphi _{1}^{\prime }-\varphi
_{2}^{\prime }\right] \psi ^{\dagger }\psi \right]  \nonumber \\
&&{\it T}\exp \left[ -i\frac{mc}{\hslash }\sqrt{2\pi G}\int dtdx\left[
\varphi _{1}-\varphi _{2}\right] \psi ^{\dagger }\psi \right] \left| \Phi
(0)\right\rangle _{\psi }  \nonumber \\
&&{\it T}\exp \left[ -i\frac{mc}{\hslash }\sqrt{2\pi G}\int dtdx\left[
\varphi _{1}+\varphi _{2}\right] \psi ^{\dagger }\psi \right] \left| \Phi
(0)\right\rangle _{\psi }  \nonumber \\
&&_{\psi }\left\langle \Phi (0)\right| {\it T}^{-1}\exp \left[ i\frac{mc}{%
\hslash }\sqrt{2\pi G}\int dtdx\left[ \varphi _{1}^{\prime }+\varphi
_{2}^{\prime }\right] \psi ^{\dagger }\psi \right]  \label{alternative}
\end{eqnarray}
where, due to the constraint on the meta-state space, $\tilde{\psi}$
operators were replaced by $\psi $ operators, and the meta-state vector $%
\left| \Phi (0)\right\rangle _{\tilde{\psi}}$ by $\left| \Phi
(0)\right\rangle _{\psi }$. This expression can even be taken as an
independent equivalent definition of the non-unitary dynamics, free from any
reference to the extended algebra including unobservable degrees of freedom.

Consider an initial linear, for simplicity orthogonal, superposition of $N$
localized states of a macroscopic body, existing, as shown above, as pure
states corresponding to unentangled bound meta-states for bodies of ordinary
density and a mass $M$ higher than $\sim 10^{11}m_{p}$\cite{defilippo3}: 
\begin{equation}
\left| \Phi (0)\right\rangle =\frac{1}{\sqrt{N}}\sum_{j=1}^{N}\left|
z_{j}\right\rangle  \label{superposition}
\end{equation}
where $\left| z\right\rangle $ represents a localized state centered in $z$.
We consider the localized states as approximate eigenstates of the particle
density operator, i.e. $\psi ^{\dagger }(x,t)\psi (x,t)\left| z\right\rangle
\simeq n(x-z)\left| z\right\rangle $, where time dependence is irrelevant,
consistently with these states being stationary both in the gravity-free and
in the interacting Schr\"{o}dinger pictures apart from a slow spreading,
which, as shown below, is much slower than the computed time for wave
function reduction.

According to Eq. (\ref{alternative}), the density matrix elements are then
given by 
\begin{eqnarray}
&&\left\langle z_{h}\right| \rho _{Ph}(t)\left| z_{k}\right\rangle  \nonumber
\\
&=&\int {\it D}\left[ \varphi _{1},\varphi _{2},\varphi _{1}^{\prime
},\varphi _{2}^{\prime }\right] \exp \frac{ic^{2}}{2\hslash }\int dtdx\left[
\varphi _{1}\nabla ^{2}\varphi _{1}-\varphi _{2}\nabla ^{2}\varphi
_{2}-\varphi _{1}^{\prime }\nabla ^{2}\varphi _{1}^{\prime }+\varphi
_{2}^{\prime }\nabla ^{2}\varphi _{2}^{\prime }\right]  \nonumber \\
&&\frac{1}{N^{2}}\sum_{j=1}^{N}\exp \left[ -i\frac{mc}{\hslash }\sqrt{2\pi G}%
\int dtdx\left[ \left[ \varphi _{1}-\varphi _{2}\right] n(x-z_{j})-\left[
\varphi _{1}^{\prime }-\varphi _{2}^{\prime }\right] n(x-z_{j})\right] %
\right]  \nonumber \\
&&\exp \left[ -i\frac{mc}{\hslash }\sqrt{2\pi G}\int dtdx\left[ \left[
\varphi _{1}+\varphi _{2}\right] n(x-z_{h})-\left[ \varphi _{1}^{\prime
}+\varphi _{2}^{\prime }\right] n(x-z_{k})\right] \right]
\end{eqnarray}
and, after integrating out the scalar fields, 
\begin{equation}
\left\langle z_{h}\right| \rho _{Ph}(t)\left| z_{k}\right\rangle =\frac{1}{%
N^{2}}\sum_{j=1}^{N}\exp \frac{i}{\hslash }Gm^{2}t\int dxdy\left[ \frac{%
n(x-z_{j})n(y-z_{h})}{|x-y|}-\frac{n(x-z_{j})n(y-z_{k})}{|x-y|}\right]
\label{coherences}
\end{equation}
which shows that, while diagonal elements are given by $\left\langle
z_{h}\right| \rho _{Ph}(t)\left| z_{h}\right\rangle =1/N$, the coherences,
under reasonable assumptions on the linear superposition in Eq. (\ref
{superposition}) of a large number of localized states, approximately
vanish, due to the random phases in Eq. (\ref{coherences}). This makes the
state $\rho _{Ph}(t)$, for times $t\gtrsim T_{G}\sim
10^{20}(M/m_{p})^{-5/3}\sec $, which are consistently short with respect to
the time of the entropic spreading $\sim 10^{3}\sec $, equivalent to an
ensemble of localized states: 
\begin{equation}
\rho _{Ph}(t)\simeq \frac{1}{N}\sum_{j=1}^{N}\left| z_{j}\right\rangle
\left\langle z_{j}\right| .
\end{equation}
It is worthwhile to remark that the extremely short localization time of a
macroscopic body may make its unlocalized states unobservable for all
practical purposes. The above analysis is also supported by numerical
evidence independently from the particular assumptions made here on the
initial unlocalized state (\ref{superposition}) \cite
{defilippomaimonerobustelli}. In such a way one gets a gravity-induced
dynamical reduction of the wave function, which up to now was assumed to
follow, possibly, from a future theory of quantum gravity\cite{karolyhazy}.
It is worthwhile to remark that the order of magnitude of decoherence times
in Eq. (\ref{coherences}) agrees with the one obtained by previous
numerological arguments for gravity-induced localization\cite{penrose}: ''
Although a detailed estimate of $T_{G}$ would require a full theory of
quantum gravity... it is reasonable to expect that for non-relativistic
systems ...''\cite{melko}. \ \ What is new here in this regard is a fully
defined dynamical model without any free parameter, which in principle
allows for the explicit evaluation of any physically relevant quantity and
for addressing crucial questions like the search for
(gravitational-)decoherence free states of the physical operator algebra\cite
{defilippo0}.

To be more specific, we have derived the first unified model for Newtonian
gravity and gravity-induced decoherence. If the states $\left|
z_{j}\right\rangle $ in Eq. (\ref{superposition}) are the pointer states of
a measurement apparatus and $\left| e_{j}\right\rangle $ are the measurement
eigenstates of a microscopic system, the product state 
\begin{equation}
\left| z_{0}\right\rangle \otimes \sum_{j}c_{j}\left| e_{j}\right\rangle
\end{equation}
according to the traditional von Neumann model for the interaction between
the two systems, is transformed into an entangled state\cite{vonneumann} 
\begin{equation}
\sum_{j}c_{j}\left| z_{j}\right\rangle \otimes \left| e_{j}\right\rangle .
\end{equation}

Obviously our previous analysis of the effect of the gravitational
(self-)interaction on the quantum motion of the macroscopic body is not
affected by the presence of the microscopic system, by which the reduction
of the wave function occurs: 
\begin{equation}
\sum_{j,k}c_{j}\bar{c}_{k}\left| z_{j}\right\rangle \otimes \left|
e_{j}\right\rangle \left\langle e_{k}\right| \otimes \left\langle
z_{k}\right| \longrightarrow \sum_{j}\left| c_{j}\right| ^{2}\left|
z_{j}\right\rangle \otimes \left| e_{j}\right\rangle \left\langle
e_{j}\right| \otimes \left\langle z_{j}\right| .  \label{reduction}
\end{equation}

Of course one can look in principle for a collapse model\cite
{ghirardi,pearle} in terms of a stochastic dynamics for pure states, which,
when averaged, leads to Eq. (\ref{reduction}). Apart, in principle, from the
non uniqueness of the stochastic realization\cite{pearle}, stochastic models
can certainly be useful as computational tools\cite{carmichael}. However the
view advocated here considers density matrices arising from gravitational
decoherence as the fundamental characterization of the system state and not
just as a bookkeeping tool for statistical uncertainties. The fact that the
apparent uniqueness of the measurement result seems to imply a real collapse
is perhaps more an ontological than a physical problem, and presumably, if
one likes it, that can be addressed by a variant of the many-world
interpretation\cite{everett,gellmann}.

\section{Black hole heuristic}

Our first aim is to evaluate within our model the finite linear dimension of
a collapsed matter lump, replacing the classical singularity. In order to do
that we boldly use Eq. (\ref{newtonlimit}) for lengths smaller than $\mu
_{0} $ and $\mu _{2}$, namely in the limit $\mu _{0},\mu _{2}\rightarrow 0$.
This corresponds to the replacement of our meta-Hamiltonian with the model
meta-Hamiltonian in Ref. \cite{defilippo1}, where there is no gravitational
interaction within observable and within hidden matter, while there is a
Newton interaction between observable and hidden matter. This interaction is
effective in lowering the gravitational energy of a matter lump as far as
the localization length $\Lambda \sim (\hslash ^{2}\Xi ^{3}/GM^{3})^{1/4}$
is fairly smaller than the lump radius $\Xi $. The highest possible density
then corresponds roughly to $(\hslash ^{2}\Xi ^{3}/(GM^{3}))^{1/4}=\Xi $ ,
namely to 
\begin{equation}
\Xi =\frac{\hslash ^{2}}{GM^{3}}.  \label{minimal}
\end{equation}

As to the space-time geometry, the Schwarzschild metric in ingoing
Eddington-Finklestein coordinates ($v,r,\theta ,\phi $) covers the two
regions of the Kruskal maximal extension that are relevant to gravitational
collapses\cite{hawkingelllis}: 
\begin{equation}
ds^{2}=-\left[ 1-2MG/\left( rc^{2}\right) \right] dv^{2}+2drdv+r^{2}\left[
d\theta ^{2}+\sin ^{2}\theta d\phi ^{2}\right] .
\end{equation}
If in the region beyond the horizon we put $x=v-\int dr\left[ 1-2MG/\left(
rc^{2}\right) \right] ^{-1}$, then 
\begin{equation}
ds^{2}=\left[ 1-2MG/\left( rc^{2}\right) \right] ^{-1}dr^{2}-\left[
1-2MG/\left( rc^{2}\right) \right] dx^{2}+r^{2}\left[ d\theta ^{2}+\sin
^{2}\theta d\phi ^{2}\right]  \label{xrmetric}
\end{equation}

If we trust (\ref{minimal}) as the minimal length involved in the collapse,
a future full theory of quantum gravity should include a mechanism avoiding
the singularity at $r=0$ by the introduction of $\Xi $ as a regularization
length. In particular, to characterize the region occupied by the collapsed
lump, consider that for time-like geodesics at constant $\theta $ and $\phi $
one can show that $\left| dx/dr\right| \sim r^{3/2}$ \ as $r\rightarrow 0$.
This implies that the $x$ coordinate difference $\Delta x$\ of two material
points has a well defined limit as $r\rightarrow 0$, by which it is natural
to assume that the $x$ width of the collapsed matter lump is $\Delta x\sim
\Xi $. As to the apparent inconsistency of matter occupying just a finite $%
\Delta x$ interval with $\partial /\partial x$ being a Killing vector, one
should expect on trans-Planckian scales substantial quantum corrections to
the Einstein equations that the model gives on a classical level, with the
dilaton and the ghost fields, though vanishing in the average, playing a
crucial role. On the other hand we are proceeding according to the usual
assumption, or fiction, of QM on the existence of a global time variable, at
least in the region swept by the lump. In fact the most natural way to
regularize (\ref{xrmetric}) is to consider it as an approximation for $r>\Xi 
$ of a regular metric, whose coefficients for $r\rightarrow 0$ correspond to
the ones in (\ref{xrmetric}) with $r=\Xi $, in which case there is no
obstruction in extending the metric to $r<0$, where taking constant
coefficients makes $\partial /\partial r$ a time-like Killing vector. As a
consequence, the relevant space metric in the region swept by the collapsed
lump is 
\[
ds_{SPACE}^{2}\sim 2MG/\left( \Xi c^{2}\right) dx^{2}+\Xi ^{2}\left[ d\theta
^{2}+\sin ^{2}\theta d\phi ^{2}\right] . 
\]
The volume of the collapsed matter lump is then: 
\begin{equation}
V\sim \Xi ^{2}\Delta x\sqrt{MG/\left( \Xi c^{2}\right) }=\left[ \hslash
^{2}/\left( GM^{3}\right) \right] ^{5/2}\sqrt{MG/c^{2}}=\hslash
^{5}M^{-7}/\left( G^{2}c\right) .
\end{equation}
According to the above view, thermodynamical equilibrium is reached, due to
the gravitational interaction generating entanglement between the observable
and hidden meta-matter, by which the matter state is a microcanonical
ensemble corresponding to the energy 
\begin{equation}
E=Mc^{2}+GM^{2}/\Xi =Mc^{2}+GM^{2}\left[ GM^{3}/\hslash ^{2}\right] \sim
G^{2}M^{5}/\hslash ^{2},\ \ if\ \ M\gg M_{P},  \label{graven}
\end{equation}
where $M_{P}=\sqrt{\hslash c/G}$ is the Planck mass, and to the energy
density 
\begin{equation}
\varepsilon =E/V\sim G^{4}cM^{12}/\hslash ^{7}.
\end{equation}
For simplicity we treat the collapsed lump as a three-dimensional bulk,
since treating it more properly, for the presence of the huge dilation
factor in the $x$ direction, as a string-like structure gives unchanged
results. As this energy density corresponds to a very high temperature, not
to be mistaken for the Hawking temperature, the matter can be represented by
massless fields, whose equilibrium entropy is given by 
\begin{equation}
S\sim \left( K_{B}/\not{h}^{3/4}c^{3/4}\right) \varepsilon
^{3/4}V=GM^{2}K_{B}/\left( \hslash c\right) .  \label{entropy}
\end{equation}
Of course this result can be trusted at most for its order of magnitude, the
uncertainty in the number of species being just one part of an unknown
numerical factor. With this proviso, common to other approaches\cite{wald},
Eq. (\ref{entropy}) agrees with B-H entropy.

Our heuristic assumption of taking as gravitational energy of the collapsed
lump just the expression given above in Eq. (\ref{graven}) is consistent
with the connection existing between the temperature of the collapsed lump
and Hawking temperature on purely thermodynamical grounds. In fact, if we
take for granted that a future theory of quantum gravity will account for
black hole evaporation, we can connect the temperature 
\begin{equation}
T\sim \sqrt[4]{\varepsilon h^{3}c^{3}}/K_{B}\sim cGM^{3}/K_{B}\hslash
\end{equation}
of our collapsed matter lump with the (spectral) temperature of the
radiation at infinity. If we model radiation by massless fields, emitted for
simplicity at a constant temperature as we are interested just in orders of
magnitude, this temperature is defined in terms of the ratio $E_{\infty
}/S_{\infty }$ of its energy $E_{\infty }$ and its entropy $S_{\infty }$. It
is natural to assume that, ''once'' thermodynamical equilibrium is reached
due to the highly non-unitary dynamics close to the classical singularity,
no entropy production occurs during evaporation, by which $S_{\infty }=S$.
Then, if $E_{\infty }=Mc^{2}$ is the energy of the total Hawking radiation
spread over a very large space volume, its temperature agrees with Hawking
temperature, i.e. \ 
\begin{equation}
T_{\infty }=\left( E_{\infty }/E\right) T\sim \left( c^{3}\hslash
/MGK_{B}\right) .
\end{equation}

\section{Concluding remarks}

Of course the reversibility of the unitary meta-dynamics makes entropy
decrease conceivable too\cite{schulman}, so that a derivation of the
entropy-growth for a closed system (in principle the whole universe), in the
present context, must have recourse to the choice of suitable initial
conditions, like unentanglement between the observable and the hidden
algebras. While the assumption of special initial conditions dates back to
Boltzmann, only a non-unitary dynamics makes it a viable starting point,
within a quantum context, for the microscopic derivation of the second law
of thermodynamics, in terms of von Neumann entropy, for a genuinely closed
system. This is meant without introducing generalized microcanocity
conditions, and then renouncing isolation\cite{gemmer}.

It should be remarked that, for a realistic physical setting, most of the in
principle observable degrees of freedom are yet out of our control and
non-unitarity is the result of interactions with both fundamentally hidden
degrees of freedom and with the environment. Environment-induced decoherence 
\cite{paz}, in most cases, may overshadow fundamental decoherence, even
though the recent amazing experimental achievements in preserving and
measuring quantum coherences make the detection of gravity-induced
decoherence a less despairing task\cite{rae,arndt,mooij,friedman,julsgaard}.
In this respect the most natural experimental setting to look for
gravitationally-induced decoherence seems to be that of Bose-Einstein (B-E)
condensation, due to the unprecedented scale of controlled quantum coherence
achieved there\cite{leggett}. In particular the localization mass-threshold
is not too far from the present experimental limits, while its robustness
with respect to the mass density variations may be a typical signature of
our gravitational self-interaction.

In conclusion, if we define a suitable non-unitary modification of fourth
order gravity, by doubling the matter algebra and introducing a suitable
constraint in order not to enlarge the observable algebra, we get the
following outcomes:

1)classical runaway solutions are absent and the ensuing classically stable
theory may be made equivalent to Einstein gravity;

2)the Newtonian limit is classically equivalent to ordinary Newton gravity;

3)from a quantum viewpoint this non-unitary limit implies gravity induced
localization and decoherence, which are compatible both with the wavelike
behavior of microscopic particles and the classicality of the center of mass
motion of macroscopic bodies;

4)the model strongly supports the interpretation of the thermodynamic
entropy of a closed system as von Neumann entropy and paves the way for the
quantum foundations of the second law of thermodynamics;

5)a bold use of the action at a distance limit of the model together with
some geometric insight coming from Einstein gravity allows us to ascribe to
the smoothed singularity of a black hole a finite entropy, which apart from
an undetermined numerical factor coincides with the B-H entropy, and a very
high temperature that is compatible with the much lower Hawking evaporation
temperature.

Acknowledgments - Financial support from M.U.R.S.T., Italy, I.N.F.M. and
I.N.F.N., Salerno is acknowledged.

\end{document}